\documentstyle[prd,aps,floats]{revtex}

\begin{document}
\preprint{SUSSEX-AST 97/10-2, gr-qc/9710048}
\draft

%
% Remove this and closure after abstract, plus preprint number,
% in electronic submission
%

\input epsf
\renewcommand{\topfraction}{1.0}
\twocolumn[\hsize\textwidth\columnwidth\hsize\csname 
@twocolumnfalse\endcsname

\title{Gravitational memory of boson stars}

\author{Diego F.~Torres\cite{dtaddress}}
 
\address{Departamento de F\'{\i}sica,  Universidad Nacional 
de La Plata, C.C. 67, 1900 La Plata, Argentina}

\author{Andrew R.~Liddle\cite{aladdress} and Franz 
E.~Schunck\cite{fsaddress}}

\address{Astronomy Centre, University of Sussex,
Falmer, Brighton BN1 9QJ, United Kingdom}

\date{\today} 
\maketitle

\begin{abstract}
`Gravitational memory' refers to the possibility that, in 
cosmologies with a time-varying gravitational `constant', 
objects such as black holes may retain a memory of conditions 
at the time of their birth. We consider this 
phenomenon in a different physical scenario, where the objects under 
consideration are boson stars. We construct boson star solutions in 
scalar--tensor gravity theories, and consider their dependence on the 
asymptotic value of the gravitational strength. We then discuss
several possible physical interpretations, including the concept of pure 
gravitational stellar evolution.
\end{abstract}

\pacs{PACS numbers: 04.40.Dg, 04.50.+h \hspace*{0.55cm} Sussex preprint 
SUSSEX-AST 97/10-2, gr-qc/9710048}

\vskip2pc]

%%%%%%%%%%%%%%%%%%%%%%%%%%%%%%%%%%%%%%%%%%%%%%%%%%%%%%%%%%%%%%%%%%%%%%%%
\section{Introduction}

A few years ago, Barrow introduced the concept of gravitational
memory \cite{Grav-Mem} by posing the following problem: what happens
to black holes during the subsequent evolution of the universe if the 
gravitational coupling $G$ evolves with time? He envisaged two possible 
scenarios. In the first, dubbed scenario A, the black hole evolves 
quasi-statically, in order to adjust its size with the changing $G$. If 
true, this means that there are no static black holes, even classically, 
during any period in which $G$ changes. In the alternative 
possibility, scenario B, the local value of $G$ within the black hole is 
preserved, 
while the asymptotic value evolves with a cosmological rate. This would mean 
that the black hole keeps a {\it memory} of the strength of gravity at the 
moment of its formation. Further analysis of the striking phenomena which 
arise in both of these scenarios was made in Ref.~\cite{BARROW-CARR}.
Black holes are of particular interest in this context, because primordial 
black holes may have formed in the very early stages of the Universe. At 
those times no direct evidence of the strength of gravity is available; 
nucleosynthesis is the earliest epoch at which significant constraints 
apply. Gravitational memory may provide a unique probe of these 
early epochs. However, in this paper our aim is to consider the 
gravitational 
memory phenomenon in a completely stellar arena. 
The motivation is twofold. On one side, 
since there are no singularities or event horizons as
in the black hole case, there are many 
calculational advantages which may provide a more direct route towards 
shedding light on which of the scenarios stated above will happen in 
practice. On the other, a comprehensive understanding of 
the effects of a varying-$G$ cosmology upon astrophysical objects is
far from complete.

To make a proper study of a varying gravitational strength, it is imperative 
to operate within a self-consistent framework rather than simply writing in 
a variation `by hand'. The most useful such framework is scalar--tensor 
theories \cite{11-BOSON}, which have an action
\begin{equation}
\label{action}
S = \int \frac{\sqrt{-g}}{16\pi} \, dx^4\left[ \phi R-
	\frac{\omega(\phi )}{\phi} \, \partial_\mu \phi \,
	\partial^\mu \phi + 16 \pi {\cal L}_{{\rm m}} \right] \,.
\end{equation}
Here $g_{\mu\nu}$ is the metric, $R$ is the scalar curvature, $\phi$ is the 
Brans--Dicke field, ${\cal L}_{{\rm m}}$ is the Lagrangian of the matter 
content of the system, and $\omega(\phi)$ gives the strength 
of the coupling between the scalar $\phi$ and the metric.
The special case of a constant $\omega$ is the Jordan--Brans--Dicke (JBD)
theory, and general relativity is regained in the limit of large $\omega$.
The gravitational constant $G$ of the Einstein--Hilbert action is replaced 
by a dynamical field, $\phi^{-1}$. In the usual 
applications, it is either spatially-constant but time-varying (cosmological 
solutions) or spatially-varying but time-independent (astrophysical 
solutions). Our situation will be unusual in including both types of 
variation. 

Our choice of stellar object is motivated by two concerns. Firstly, it 
should be simple enough as to allow us to isolate the effects caused by a 
varying $G$. Secondly, in order to show the changes due to the
particular gravitational theory, it should be fully relativistic, 
in the sense that its equilibrium configuration must be obtained from 
Einstein-like field equations. 
There is one type of stellar object which meets these criteria, though so 
far it is known to exist only as a theoretical construct. 
It is the boson star, the analogue of a neutron star 
formed when a large collection of bosonic particles (normally taken 
to be scalars) becomes gravitationally bound. Such configurations were 
introduced by Kaup \cite{Kaup} and by Ruffini and Bonazzola \cite{RB} in 
the nineteen sixties, but current interest was sparked by Colpi et 
al.~\cite{CSW} who showed that, provided the scalar field has a 
self-interaction, the boson star masses could be of the same order of 
magnitude as, or even much greater than, the Chandrasekhar mass.
This led to the study of many properties, which are summarized 
in two reviews~\cite{reviews}. It seems possible for them to form through 
gravitational collapse of a scalar field \cite{MLT}, though little is known 
as yet. The observational status of boson
stars was analyzed very recently in Ref.~\cite{LIDDLE-SCHUNCK}, 
where it was asked whether radiating baryonic matter moving around 
a boson star could be converted into an observational signal. Unfortunately, 
any direct detection looks a long way off.

As far as we are aware, only three papers have studied boson stars in 
scalar--tensor theories. Gunderson and Jensen \cite{GUNDERSON} studied the 
JBD scenario, concluding that typically the mass of 
equilibrium solutions would be reduced by a few percent. This work was 
generalized by Torres \cite{TORRES_B}, who looked at several coupling 
functions $\omega(\phi)$, chosen to be compatible with known weak-field 
limit and nucleosynthesis constraints. Similar scenarios were further 
studied in Ref.~\cite{CS}. In this paper we shall study boson stars
formed at different times of cosmic history 
and in addition consider the possible physical evolution of such objects.
One might further hope that results found for boson 
stars might be representative of other stellar candidates, 
especially neutron stars but perhaps even indicative of what 
might happen with long-lived hydrogen-burning stars such 
as our own Sun. 

\section{Equilibrium configurations}

We begin with a review of the formalism, which can be found in more detail 
in Ref.~\cite{TORRES_B}. The material from which the boson star is made is a 
complex, massive, self-interacting scalar field $\psi$, which is unrelated 
to the Brans--Dicke scalar already described. Its Lagrangian is
\begin{equation}
{\cal L}_{{\rm m}} = -\frac{1}{2} g^{\mu \nu} \, \partial_\mu \psi^*
	\partial_\nu \psi -\frac{1}{2} m^2 |\psi|^2 
	-\frac{1}{4} \lambda |\psi|^4 \,.
\end{equation}
The $U(1)$ symmetry leads to conservation of boson number. Varying the 
action with respect to $g^{\mu\nu}$ and $\phi$ we obtain the field 
equations:
\begin{eqnarray}
\label{field0}
R_{\mu\nu}-\frac{1}{2} g_{\mu\nu}R & = & \frac{8\pi}{\phi}T_{\mu\nu}
	+\frac{\omega(\phi )}{\phi} \left( \phi_{,\mu} \phi_{,\nu}-
	\frac{1}{2} g_{\mu \nu }\phi^{,\alpha}\phi_{,\alpha}\right)
	\nonumber \\
 & & \quad + \frac{1}{\phi} \left( \phi_{,\mu;\nu} -g_{\mu\nu} \Box\phi
	\right) \,, \\
 \label{field00}
\Box \phi & = & \frac{1}{2\omega+3} \left[ 8\pi T-\frac{d\omega}{d\phi}
	\phi^{,\alpha}\phi_{,\alpha}\right] \,,
\end{eqnarray}
where $T_{\mu\nu}$ is the energy--momentum tensor for the
matter fields (Eq.~(5) of Ref.~\cite{TORRES_B})
and $T$, its trace. Commas and semicolons are 
derivatives and covariant derivatives, respectively. 

At first, we seek static equilibrium solutions. The metric of a 
spherically-symmetric system can be written as 
\begin{equation}
\label{metric}
ds^2=-B(r) dt^2 + A(r) dr^2 +r^2 d\Omega^2 \,.
\end{equation}
We also demand a spherically-symmetric form for the $\psi$ field. Crucially, 
the most general ansatz consistent with the static metric 
permits a time-dependent $\psi$ of the form
\begin{equation}
\label{boson}
\psi(r,t)=\chi(r) \exp{[-i\varpi t]} \,.
\end{equation}
To write the equations of structure of the star, we use a rescaled radial 
coordinate, given by
\begin{equation}
\label{x}
x=mr \,.
\end{equation}
{}From now on, a prime will denote a derivative with respect to the variable
$x$. We also define dimensionless quantities by
\begin{equation}
\label{dimensionless}
\Omega=\frac{\varpi}{m}\; , \; \; \Phi=\frac{\phi}{m_{{\rm Pl}}^2} \; ,\; \; 
  \sigma=\sqrt{4\pi} \frac{\chi(r)}{m_{{\rm Pl}}} \; ,\; \;  
  \Lambda=\frac{\lambda}{4\pi} \left( \frac{m_{{\rm  Pl}}}{m} \right)^2,
\end{equation}
where $m_{{\rm Pl}} \equiv G_0^{-1/2}$ is the present Planck mass. Our 
observed gravitational coupling implies $\Phi = 1$. In order to 
consider the total amount of mass of the star within a radius $x$ we change 
the function $A$ in the metric to its Schwarzschild form, 
\begin{equation}
\label{M}
A(x)=\left(1-\frac{2M(x)}{x\,\Phi(\infty)}\right)^{-1}.
\end{equation}
The issue of the definition of mass in JBD theory is quite a subtle one 
\cite{Whinnett}. The above gives the Schwarzschild mass; there are arguments 
that the tensor mass is more appropriate \cite{Whinnett}, but we have verified 
numerically that for the large couplings we use the difference is negligible.

Note that a factor $\Phi(\infty)$ appears in Eq.~(\ref{M}). This
is crucial to obtain the correct value of the mass, which is given by
\begin{equation}
M_{{\rm star}}= M(\infty) \,\Phi(\infty)\, \frac{m_{{\rm Pl}}^2}{m}\,,
\end{equation}
for a given value of $m$. The $\Phi(\infty)$ factor allows for the asymptotic 
gravitational coupling to be different from that presently 
observed.\footnote{Note that the $\Phi(\infty)$ factor was mistakenly 
forgotten in Ref.~\cite{TORRES_B}.
This affects Table III of that work, where $M(\infty)$ is actually
$M(\infty)/\Phi(\infty)$. The $\Phi(0)$ values and the 
conclusions extracted from it are unaffected.}
With all these definitions, 
the non-trivial equations of structure are \cite{TORRES_B}
\begin{eqnarray}                     
\label{field1}
\sigma^{\prime \prime} & + & \sigma^{\prime} \left( \frac{B^\prime}{2B} -
	\frac{A^\prime}{2A} + \frac{2}{x} \right) \nonumber \\
  && + A \left[ \left(\frac{\Omega^2}{B}-1 \right)\sigma - 
	\Lambda \sigma^3 \right]=0 \; , 
\end{eqnarray}
\begin{eqnarray}
\label{field2}
\Phi^{\prime \prime} & + & \Phi^{\prime} \left( \frac{B^\prime}{2B} -
	\frac{A^\prime}{2A} + \frac{2}{x} \right)+
	\frac{1}{2\omega+3}\frac{d\omega}{d\Phi} \Phi^{\prime 2}
	\nonumber \\
 && - \frac{2A}{2\omega+3} \left[ \left(
	\frac{\Omega^2}{B}-2 \right)\sigma^2 -\frac{\sigma^{\prime 2}}{A} - 
	\Lambda \sigma^4 \right] =0 \; , 
\end{eqnarray} 
\begin{eqnarray}
\label{field3}
\frac{B^{\prime}}{xB} & - & \frac{A}{x^2}\left( 1-\frac{1}{A} \right)=
	\frac{A}{\Phi} \left[ \left( \frac{\Omega^2}{B}-1 \right)
	\sigma^2 +\frac{\sigma^{\prime 2}}{A} - 
	\frac{\Lambda}{2} \sigma^4 \right] \nonumber \\
 && + \frac{\omega}{2}\left(\frac{\Phi^{\prime}}{\Phi}\right)^2
	+ \left( \frac{\Phi^{\prime \prime}}{\Phi}-
	\frac{1}{2}\frac{\Phi^{\prime}}{\Phi} \frac{A^\prime}{A} 
	\right) + \frac{1}{2\omega+3}\frac{d\omega}{d\Phi}
	\frac{\Phi^{\prime 2}}{\Phi} \nonumber \\
 &&  - \frac{A}{\Phi} \frac{2}{2\omega+3} \left[ 
	\left(\frac{\Omega^2}{B}-2 \right)\sigma^2
	-\frac{\sigma^{\prime 2}}{A} - \Lambda \sigma^4 \right] \; , 
\end{eqnarray}
\begin{eqnarray}
\label{field4}
\frac{2BM^\prime}{x^2 \Phi(\infty)} & = & \frac{B}{\Phi} \left[ \left(
	\frac{\Omega^2}{B}+1 \right)\sigma^2 +\frac{\sigma^{\prime 2}}{A}
	+ \frac{\Lambda}{2} \sigma^4 \right] + \frac{\omega}{2}\frac{B}{A}
	\left( \frac{\Phi^{\prime}}{\Phi}\right)^2\nonumber \\
 && + \frac{B}{\Phi} \frac{2}{2\omega+3} \left[
	\left(\frac{\Omega^2}{B}-2 \right)\sigma^2
	-\frac{\sigma^{\prime 2}}{A} - \Lambda \sigma^4 \right] \nonumber \\
 && -\frac{B}{A(2\omega+3)}\frac{d\omega}{d\Phi} 
	\frac{\Phi^{\prime 2}}{\Phi} - \frac{1}{2}
	\frac{\Phi^{\prime}}{\Phi} \frac{B^\prime}{A} \,.
\end{eqnarray}
To solve these equations numerically, we use a fourth-order Runge--Kutta
method, for which details may be found in Ref.~\cite{TORRES_B}. Here we 
shall study two kinds of theories. 

\begin{enumerate}
\item JBD theory with $\omega=400$. This is comparable to current observational 
limits \cite{reas,jbdnuc}.
\item A scalar--tensor theory with a coupling function of the form 
$$2\omega+3=2B_1 |1-\Phi|^{-\alpha} \,,$$
where we choose $\alpha = 0.5$ and $B_1=5$. The cosmological setting of 
this model has been studied by Barrow and 
Parsons \cite{BARROW-PARSONS}.
\end{enumerate}
As explained in Ref.~\cite{TORRES_B}, the results for these couplings
may be thought of as the general behavior of any scalar--tensor
theory, by suitable expanding a general coupling in a Taylor or Laurent 
series. The positive exponent of this theory must be obtained from power-law
couplings, but the behavior of these was found to be similar to the
pure Brans--Dicke ones.

In addition to the freedom to choose the fundamental parameters, there are 
two free boundary conditions. One is the value of the boson field at the 
centre of the star, $\sigma(0)$, which we call the central density. The 
second is the asymptotic value of the Brans--Dicke field, which determines 
the asymptotic strength of gravity. The other boundary conditions are fixed 
by demanding non-singularity and finite mass \cite{reviews,TORRES_B}. This 
still leaves an infinite discrete set of solutions with different $\varpi$, 
corresponding to a different number of nodes in $\sigma(x)$. We choose the 
nodeless solution, which is the only stable one. Higher node solutions are 
generated in Ref.~\cite{CS}.

\section{Quasi-static evolution}

We first assess the likely cosmological variation of $\phi$, concentrating on 
JBD theory. During radiation domination, the attractor behavior is actually 
exactly that of general relativity, namely a constant $\phi$ and $a \propto 
t^{1/2}$. This changes with the onset of matter domination, when the 
attractor solution becomes \cite{Nariai,Gurevich} 
\begin{equation}
a(t) \propto t^{(2-n)/3} \quad ; \quad G(t) \propto t^{-n} \,,
\end{equation}
where $n=2/(4+3\omega)$, so $G$ exhibits a slow decrease. Assuming that 
matter--radiation equality took place near the general relativity value, at 
$z_{{\rm eq}} = 24\,000 \, \Omega_0 h^2$, then, for critical density and $h = 
0.5$, the fractional change in $G$ since equality is
\begin{equation}
\frac{G(t_0)}{G(t_{{\rm eq}})} = 6000^{-1/(1+\omega)} \,.
\end{equation}
For $\omega = 400$, the ratio is 0.98, so $G$ will have changed value by 
about two percent since equality.
 
\begin{figure}[t]
\centering 
\leavevmode\epsfysize=6cm \epsfbox{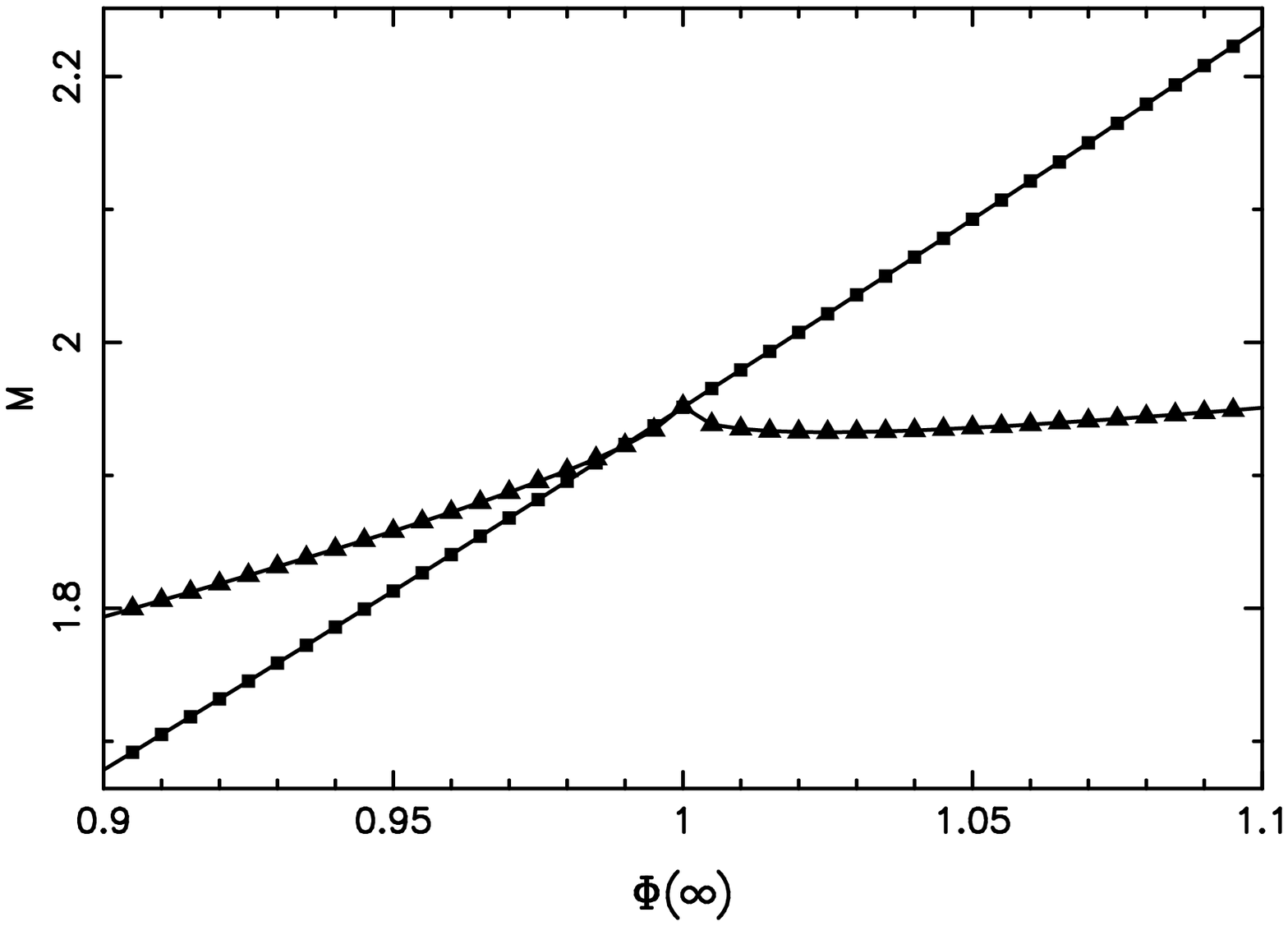}\\ 
\leavevmode\epsfysize=6cm \epsfbox{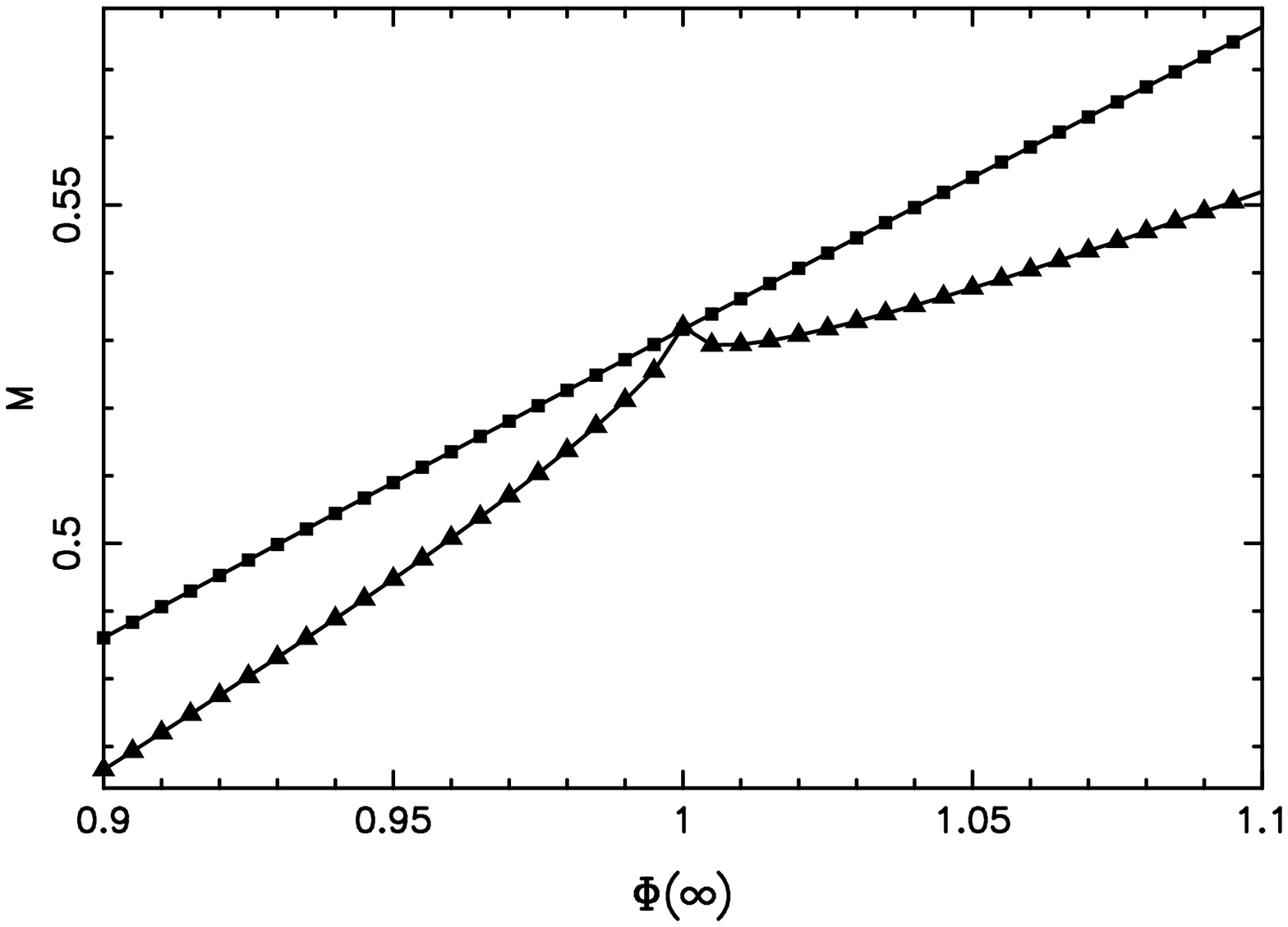}\\ 
\caption[fig1]{\label{fig1} Boson star masses as a function of 
$\Phi(\infty)$; squares are for the JBD theory and circles the 
scalar--tensor theory. The upper panel
shows models with $\Lambda=100$ and $\sigma(0)=0.06$. The lower one has
$\Lambda=0$ and $\sigma(0)=0.1$.}
\end{figure} 

Informed by the likely range of variation, we compute static configurations 
with different values of the asymptotic gravitational strength, denoted 
$\Phi(\infty)=1/G$. In Fig.~1, we see that if $\Lambda$ and the central 
density $\sigma(0)$ are kept fixed, then the mass changes with a different 
$\Phi(\infty)$. In the JBD case, the mass is a growing function of 
$\Phi(\infty)$, while in the more complicated second model there 
is actually a small peak in the mass at our present gravitational strength. 
That peak is due to the special form of the coupling function, 
which selects out our own observed coupling strength. 

Let us now suppose that evolution does drive the system through a series of 
quasi-static states, and na\"{\i}vely estimate the sorts of energies 
involved. Fig.~1 shows that variation of $\Phi(\infty)$ by a few percent can 
change the masses of configurations by a few percent. As it happens, this is 
very similar to, or even greater than, the fraction of the mass--energy 
which can be liberated by nuclear reactions over the lifetime of a star, a 
process which for most stars occurs on similar timescales to the cosmological 
ones of the time-varying $G$.

\begin{figure}[t]
\centering 
\leavevmode\epsfysize=6cm \epsfbox{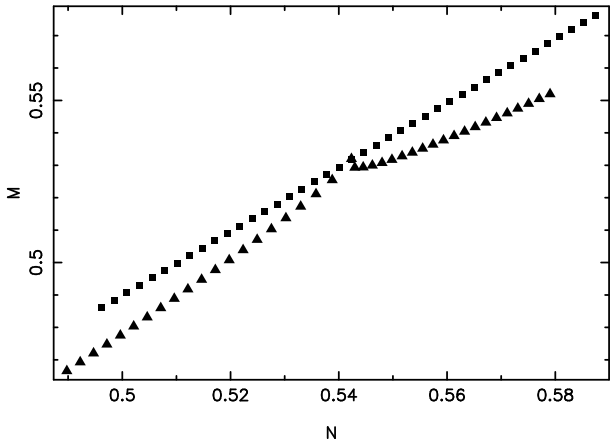}\\ 
\caption[fig2]{\label{fig2} This shows the boson mass as a function of 
particle number in the two models, with squares for the JBD 
theory and triangles for the scalar--tensor one. Both have $\Lambda = 0$ and 
$\sigma(0) = 0.1$. The leftmost point in each case corresponds to 
$\Phi(\infty) 
= 0.9$ and the rightmost to $\Phi(\infty) = 1.1$. As $M(N)$ is single valued, 
there can be no change of stability.}
\end{figure} 

Before analyzing this in further detail, we shall first show the 
computation for the number density of the configurations, defined by
\begin{equation}
\frac{m^2}{m_{{\rm Pl}}^2} N= \Omega \int_0^\infty  \sigma^2 
	\sqrt{\frac AB} \, x^2 \, dx \,,
\end{equation} 
We found that, for both models under 
consideration, the number of bosons is 
a growing function of $\Phi(\infty)$.
Fig.~2 shows the bifurcation diagram, number density against mass.
There is no cusp structure in the sense of catastrophe theory,
which implies that there is no change in the stability criterion for values 
of $G$ close to the present one~\cite{KUSMARTSEV}. In fact, we found this 
result 
for a wider range of $\Phi(\infty)$ than shown.
 
\section{Scenarios}

Now we discuss possible interpretations of the computational results 
described above, in the same context which led Barrow to propose the 
gravitational memory hypothesis. We describe the possible subsequent 
histories a boson star might have, after forming at a time $t_{{\rm f}}$ 
with a mass $M(t_{{\rm f}})$.

\subsection{The gravitational memory hypothesis}

The star remains completely static, without change in the values 
of its mass or central density. Such a situation would be reminiscent say of 
a virialized galaxy after gravitational collapse, which has become decoupled 
from the cosmological expansion of the Universe around it and in particular 
does not participate in further expansion. In such a scenario the mass
$M(t_{{\rm f}})$ is a function of the formation time, and hence of
$\Phi(\infty\,,\,t_{{\rm f}})$, as well as the central density; the star 
keeps 
memory of the value of $G$ at formation. This situation clearly cannot be 
precisely correct since the asymptotic gravitational constant does evolve, 
but it may well be that the effect on the star is much slower even than the 
cosmological evolution of $G$, so that in practice the star can be viewed as 
static.

In this scenario, there is the interesting feature that stars of the same 
mass may differ in other physical properties (their radius, for example) 
depending on the formation time.

\subsection{Quasi-static evolution hypothesis}

The opposite regime has the adjustment time for the star 
being much shorter than the asymptotic evolution of $G$. 
If true, the star should quasi-statically 
evolve, either changing its mass or its central density or both. In this 
scheme, when intervals of time are short enough compared with the scale of 
cosmological evolution, the star can be taken as static and the solutions 
computed hold in this limit. 

An interesting subcase is {\it pure gravitational evolution}. 
This assumes that the mass of the star is preserved all along
the quasi-static evolution, through the central density 
evolving in the appropriate way. This leads to 
the remarkable conclusion that stars may be able to evolve even without 
absorption or emission of energy, and without nuclear burning. Such 
quiescent evolution would be entirely gravitational in its nature. 

Purely gravitational evolution would doubtless have significant consequences 
for stellar evolution, and indeed may well be quite strongly constrained. 
Yet within the quasi-static evolution hypothesis, it is actually rather 
conservative, because the other possibility is that mass is lost during the 
evolution, if dissipative processes are required to keep tracking the 
evolutionary sequence. We have seen that the variation in $G$ can easily 
lead to changes in the mass of up to a few percent, which can be of the same 
order as the energy liberated in a conventional star by nuclear burning. 

The reduction in central density seems reasonable when we recall the force 
balance in polytropic stars. Since gravity is reducing in strength as time 
goes by, the equilibrium configurations can become more diffuse and hence 
drop in central density. In an 
extreme situation, one might wonder whether the continuation of this process 
can in fact lead to the complete destruction of the star, though this may be 
prevented by the initial negative binding energy of a stable boson star, 
since boson number is conserved.

The quasi-static evolution hypothesis has an important difference from the 
gravitational memory hypothesis; in it, stars of the same mass are identical 
in all their other physical properties too.

\subsection{Feedback on the asymptotic gravitational constant?}

Finally, one can ask whether the formation of boson stars can significantly 
influence the `asymptotic' gravitational constant; when the stars form 
gravity becomes weaker in their interior as the $\phi$ value increases. 
Following the results of Ref.~\cite{TORRES_B} (Table III), this change can 
be about 1\% between the internal and the external $G$ value. In a static 
configuration, the radius at which $\phi$ finally approaches its asymptotic 
value is quite a bit larger than the region in which the $\psi$ field is 
localized. If a very high density of boson stars formed, might they be able 
to reduce the gravitational interaction strength in quite a significant 
region around 
themselves? It is an interesting possibility, though unlikely in practice as 
the density of material available to make the stars is so small; boson stars 
would be expected to be separated by similar distances to conventional 
stars (which could also contribute to the effect). 

\section{Discussion}

In this paper, we have not directly tried to answer the question as to 
whether or not the gravitational memory phenomenon exists. Rather, we have 
introduced a new framework in which it can be studied, that of stellar 
systems. As a spinoff, it allows us to introduce the concept 
of pure gravitational stellar evolution. 
As well as their possible physical relevance, boson stars have 
the considerable advantage of being much simpler than their black hole 
analogues. They are also much simpler than neutron stars, 
as they are based directly on a field theory description. 
In that light, we have studied static configurations in two different 
scalar--tensor theories, in particular emphasizing the dependence of the 
mass on the asymptotic value of the gravitational coupling. We have also 
been able to make some preliminary statements on the dynamical stability of 
the configurations, an issue we leave for further study in a future 
publication. The next step would be numerical simulation including dynamical 
evolution, starting from a static solution and varying the asymptotic 
gravitational coupling. This looks a promising avenue for determining which 
of the proposed scenarios is the correct one, and that knowledge may allow a 
test of general relativity not just at the present epoch, but in the distant 
past, through astrophysical systems.

\acknowledgments

D.F.T. was supported by a British Council Fellowship and CONICET, A.R.L. by 
the Royal Society and F.E.S. by an European Union Marie Curie TMR 
fellowship. We thank John Barrow, Eckehard W.~Mielke, Hisa-aki Shinkai and 
Andrew Whinnett for useful discussions.

%%%%%%%%%%%%%%%%%%%%%%%%%%%%%%%%%%%%%%%%%%%%%%%%%%%%%%%%%%%%%%%%%%%%%%%%%


\begin{references}
\vspace*{-1cm}
\bibitem[\clubsuit]{dtaddress} E-mail address: {\tt
	dtorres@venus.fisica.unlp.edu.ar}
\bibitem[\diamondsuit]{aladdress} E-mail address: {\tt
	a.liddle@sussex.ac.uk}
\bibitem[\spadesuit]{fsaddress} E-mail address: {\tt
	fs@astr.cpes.susx.ac.uk}
\bibitem{Grav-Mem} J. D. Barrow, Phys. Rev. D {\bf 46}, 3227 (1992),
	Gen. Rel. Grav. {\bf 26}, 1 (1994).
\bibitem{BARROW-CARR} J. D. Barrow and B. J. Carr, Phys. Rev. D
	{\bf 54}, 3920 (1996).
\bibitem{11-BOSON} C. Brans and R. H. Dicke, Phys. Rev. {\bf 124 }, 925 
	(1961); P. G. Bergmann, Int. J. Theor. Phys. {\bf 1}, 25 (1968); 
	K. Nortvedt, Astrophys. J. {\bf 161}, 1059 (1970); R. V. Wagoner,
	Phys. Rev. D {\bf 1}, 3209 (1970).
\bibitem{Kaup} D. J. Kaup, Phys. Rev. {\bf 172}, 1331 (1968).
\bibitem{RB} R. Ruffini and S. Bonazzola, Phys. Rev. {\bf 187}, 1767 (1969).
\bibitem{CSW} M. Colpi, S. L. Shapiro and I. Wasserman, Phys. Rev. D
	{\bf 57}, 2485 (1986).
\bibitem{reviews} P. Jetzer, Phys. Rep. {\bf 220}, 163 (1992); A. R. 
	Liddle and M. S. Madsen, Int. J. Mod. Phys. {\bf D1}, 101 (1992).
\bibitem{MLT} M. S. Madsen and A. R. Liddle, Phys. Lett. B {\bf 251}, 507
	(1990); I. I. Tkachev, Phys. Lett. B {\bf 261}, 289 (1991).
\bibitem{LIDDLE-SCHUNCK} F. E. Schunck and A. Liddle, Phys. Lett. B
	{\bf 404}, 25 (1997).
\bibitem{GUNDERSON} M. A. Gunderson and L. G. Jensen, Phys. Rev. D
	{\bf 48}, 5628 (1993).
\bibitem{TORRES_B} D. F. Torres, Phys. Rev. {\bf D56}, 3478 (1997).
\bibitem{CS} G. L. Comer and H. Shinkai, Report No. gr-qc/9708071 (1997).
\bibitem{Whinnett} A. W. Whinnett, Report No. gr-qc/9711080 (1997).
\bibitem{reas} R. D. Reasenberg et al., Astrophys. J. {\bf 234}, L219 
	(1979).
\bibitem{jbdnuc} F. S. Accetta, L. M. Krauss and P. Romanelli, Phys. Lett.
	B {\bf 248}, 146 (1990); J. A. Casas, J. Garc\'{\i}a-Bellido and 
	M. Quir\'os, Phys. Lett. B {\bf 278}, 94 (1992).
\bibitem{BARROW-PARSONS} J. D. Barrow and P. Parsons, Phys. Rev. D
	{\bf 55}, 3906 (1997).
\bibitem{Nariai} H. Nariai, Prog. Theor. Phys. {\bf 42}, 544 (1969).
\bibitem{Gurevich} L. E. Gurevich, A. M. Finkelstein and V. A. Ruban,
	Astrophys. Space Sci. {\bf 98}, 101 (1973).
\bibitem{KUSMARTSEV} F. V. Kusmartsev, E. W. Mielke and F. E. Schunck,
	Phys. Rev. D {\bf 43}, 3895 (1991); Phys. Lett. A {\bf 157}, 465
	(1991).
\end{references}
\end{document}